\documentclass[fleqn,usenatbib]{mnras}

\usepackage{graphicx}	
\usepackage{amsmath}    

\newcommand{\eff}{_{\rm eff}}

\usepackage{newtxtext,newtxmath}

\usepackage[T1]{fontenc}
\usepackage{ae,aecompl}

\title[UDGs stellar kinematics]{NIHAO XXIV: Rotation or pressure supported systems? \\ Simulated Ultra Diffuse Galaxies show a broad distribution in their stellar kinematics}

\author[S. Cardona-Barrero et al.]{Salvador Cardona-Barrero$^{1,2}$
    \thanks{Contact e-mail: \href{mailto:scardona@iac.es}{scardona@iac.es}}, 
Arianna Di Cintio$^{1,2}$\thanks
    {Marie-Sk\l{}odowska-Curie Fellow}, 
Christopher B. A. Brook$^{1, 2}$\thanks
    {Ramon y Cajal Fellow},   \newauthor Tomas Ruiz-Lara$^{1,2}$,  Michael A. Beasley$^{1,2}$, Jesus Falc\'on-Barroso$^{1,2}$ \newauthor and Andrea V. Macci\`o$^{3,4,5}$ 
\\
$^{1}$ Instituto de Astrof\'isica de Canarias, Calle V\'ia L\'actea s/n, E-38206 La Laguna, Tenerife, Spain \\
$^{2}$ Universidad de La Laguna Avda. Astrof\'isico Fco. S\'anchez, E-38205 La Laguna, Tenerife, Spain \\
$^{3}$ New York University Abu Dhabi, PO Box 129188 Abu Dhabi, United Arab Emirates \\
$^{4}$ Max Planck Institute f\"ur Astronomie, K\"onigstuhl 17, D-69117 Heidelberg, Germany \\
$^{5}$ Center for Astro, Particle and Planetary Physics (CAP$^3$), New York University Abu Dhabi
}

\date{Accepted XXX. Received YYY; in original form ZZZ}

\pubyear{2019}

\begin{document}
\label{firstpage}
\pagerange{\pageref{firstpage}--\pageref{lastpage}}


\maketitle

\begin{abstract}
   In recent years a new window on galaxy evolution opened, thanks to the increasing discovery of   galaxies with a low surface brightness, such as Ultra Diffuse Galaxies (UDGs).    
   The formation mechanism of these systems is still a much debated question, and so are their kinematical properties.
   In this work, we  address  this topic by analyzing the stellar kinematics of  isolated UDGs formed in the hydrodynamical  simulation suite NIHAO. We construct projected line-of-sight velocity and velocity dispersion maps to compute the projected specific angular momentum, $\lambda_{\rm R}$, to characterize the kinematical support of the stars in these galaxies.
   We found that UDGs cover a broad distribution, ranging from dispersion to rotation supported galaxies, with similar abundances in both regimes.
   The degree of rotation support of simulated UDGs correlates with several properties such as  galaxy morphology, higher HI fractions and larger effective radii with respect to the dispersion supported group, while the dark matter halo spin and mass accretion history are similar amongst the two populations. We demonstrate that the alignment of the infalling baryons into the protogalaxy at early $z$ is the principal driver of  the $z$=0 stellar kinematic state:  pressure supported isolated UDGs form via mis-aligned gas accretion while rotation supported ones   build-up their baryons in an ordered manner. Accounting for random inclination effects, we predict that a comprehensive survey will find nearly half of field UDGs  to have rotationally supported stellar disks, when selecting UDGs with effective radius larger than $1$ kpc.
\end{abstract}

\section{Introduction}

During the latter part of the 1980s, astronomers began to find galaxies so faint and diffuse that they were barely distinguishable from the sky background \cite[e.g][]{1987AJ.....94...23B, 1985AJ.....90.2487B, 1988AJ.....95.1389S, 1988ApJ...330..634I}. The discovery of such galaxies, known as Low Surface Brightness galaxies (LSB) implied a challenge to the standard $\Lambda$CDM paradigm of galaxy formation \citep[see][]{1997PASP..109..745B}. In recent years, improvements in  instrumentation and the development of new reduction techniques  have allowed to reach much lower magnitude and surface brightness limits \citep{2014ApJ...787L..37M, 2015ApJ...798L..45V, 2016MNRAS.456.1359F, 2016ApJ...823..123T, 2010AJ....140..962M}, opening a new window on the low surface brightness universe and leading  to the discovery of 
Ultra Diffuse Galaxies (UDGs, \citealt{2015ApJ...798L..45V} see, however, \citealt{1996MNRAS.283...18D} for earlier references to galaxies with similar properties).
UDGs have been found in low redshift clusters \citep{2016A&A...590A..20V, 2018MNRAS.481.4381M} such as Virgo \citep{2005ApJ...631L..41M,2016ApJ...819L..20B},  Fornax \citep{2015ApJ...813L..15M} and the Coma cluster \citep{2015ApJ...807L...2K, 2016ApJ...830...23B, 2015ApJ...798L..45V}, in which  more than $1000$ such galaxies have been identified. Nevertheless, they have started to be found in less dense environments: \cite{2016AJ....151...96M} have found a UDG in a filament at the outskirts of the Piscis-Perseus super-cluster, \cite{2019MNRAS.488.2143P}  have used the Kilo-Degree Survey (KiDs) to look for field UDGs, and \cite{2019MNRAS.486..823R}  identified a red UDG in a void using the  IAC Stripe82 Legacy Survey \citep{2016MNRAS.456.1359F} . It is not clear if the large difference between the number of field UDGs and in-cluster UDGs is due to a mechanism the makes UDGs more likely to form in high density environments or rather it is related to an observational bias, since it is easier to infer the distance to these systems in denser regions, and thus to identify them as UDGs \citep{2017MNRAS.468..703R}. 


In general, UDG properties are  related to the environment they live in: \cite{2017MNRAS.468..703R} have studied the UDG population in Abell $168$  cluster and  find that  bluer UDGs live in the outskirts of the cluster. \cite{2017MNRAS.468.4039R} using UDGs in Hickson Compact Groups, presented a possible evolution mechanism that links blue HI rich isolated systems with the redder and HI poor UDGs that are found in clusters  \citep[see also][]{2017ApJ...836..191T}. \cite{2018ApJ...855...28S}  used the same sample and showed that these blue UDGs have huge reservoirs of HI gas.

As  suggested by their stellar and globular cluster kinematics, UDGs seem to be dark matter dominated galaxies,  which in general  follow the stellar to halo mass relation \citep[see][]{2016ApJ...819L..20B,2018ApJ...856L..31T,2019ApJ...880...91V}. 

The most astonishing property of  UDGs is that, despite having stellar masses of small dwarfs (${\rm M}_{\ast}\sim 10^{6.5-9}{\rm M}_{\odot}$), they have effective radii greater than $1\,{\rm kpc}$, similar to the ones of large spirals. These two aspects combined are the reason for which UDGs have such low effective surface brightness, reaching $27\, {\rm mag}\,{\rm arcsec}^{-2}$ \citep[e.g.][]{2016ApJS..225...11Y}.
As discussed in \cite{2020MNRAS.493...87T}, however, the effective radius is strongly dependent on a galaxy's profile, and interpreting it as the size of the galaxy should be done carefully. Indeed, they introduced a new parameter, i.e. the radius at which the stellar density decreases under $1 {\rm M}_{\odot}\, {\rm pc}^{-2}$, which seems to be more closely related to the physical size of a galaxy:  applying this parameter to UDGs, \citep{2020A&A...633L...3C} found no evidence for them to be abnormally large compared with other dwarfs.
The question that arises from their work, thus, is not about their sizes, but rather on why UDGs have such a lack of central cusp in their surface brightness profiles, unlike 
other dwarfs of similar stellar masses.

 As of today, there is no consensus amongst the scientific community regarding the formation mechanisms of UDGs. We can distinguish two groups of proposed formation mechanisms:

\begin{enumerate}
\item  the ones that require UDGs to live in high density regions, in which case  environmental effects like ram-pressure stripping at early times due to the galaxy infalling into the cluster \citep{2015MNRAS.452..937Y,2019arXiv190805684T,2020AAS...23531602T,2020MNRAS.tmp.1063S}, tidal heating \citep{2018AAS...23141205C}, or two body encounters \citep{2018NewA...60...69B} may explain the properties of in-cluster UDGs;
\item the ones that invoke internal mechanisms within the UDGs themselves, and which allow such galaxies to form also in the field. The high-spin scenario (initially proposed by \citealt{2016MNRAS.459L..51A}, see also \citealt{2017MNRAS.470.4231R,2019MNRAS.490.5182L}) suggests that UDGs may be the natural extension of the dwarf population that live into high-spin dark matter haloes, whilst the feedback-driven scenario, proposed by 
   \cite{2017MNRAS.466L...1D} and later supported by \citet{2018MNRAS.478..906C}, using the NIHAO simulation suite \citep{2015MNRAS.454...83W} and FIRE simulation \citep{2014MNRAS.445..581H} respectively, shows that successive supernovae events  generate feedback-driven outflows that flatten the density profiles of both dark matter and stars, leading to the low surface brightness characteristic of  UDGs. 
\end{enumerate}

The formation mechanisms mentioned above are not mutually exclusive and they can conspire together in order to form the whole population of UDGs (see \citealt{2019ApJ...884...79C}). 

It is challenging to understand the emergence of  UDGs in a cosmological context. In particular, it is difficult to simulate such small systems, which require high resolution to resolve sub-grid physics, within dense environments that include many more massive galaxies \citep[although see the recent work by ][]{2019arXiv190805684T,2020AAS...23531602T,2020MNRAS.tmp.1063S}. By contrast, zoom-in simulations can resolve low mass systems within less dense field environments, and such simulations have been able to reproduce populations of UDGs, as shown in  \cite{2017MNRAS.466L...1D} and \cite{2019MNRAS.487.5272J} using the NIHAO simulation suite  \citep{2015MNRAS.454...83W}, in \cite{2018MNRAS.478..906C} using the FIRE simulations \citep{2014MNRAS.445..581H} or \cite{2019MNRAS.490.5182L} using AURIGA simulations \citep{2017MNRAS.467..179G}.

Due to the diffuse nature of these systems there are only few and recent attempts to try to measure velocity gradients: \cite{2019ApJ...883L..33M} analyzed the kinematics of the HI component of $6$ UDGs, finding lower circular velocity than the expected from their baryonic Tully-Fisher relation; \cite{2019MNRAS.488.3222S} identified a velocity gradient in the HI disk of UGC 2162 compatible with what expected from dwarfs of similar masses; \cite{2017ApJ...842..133L} made a systematic analysis of the HI rich sources form the ALFALFA catalog \citep{2005AJ....130.2598G};   \cite{2018MNRAS.478.2034R} studied, via spectroscopic analysis of their stars, $5$ UDGs in the Coma Cluster providing upper limits to the stellar rotation of these systems, showing that they are compatible with being dwarfs; \cite{2019ApJ...884...79C} measured, via intermediate-resolution spectra, the kinematics of $9$ UDGs in the Coma cluster, finding signals of major axis rotation in $3$ of them;
\cite{2019A&A...625A..76E}, using  MUSE integral field spectroscopy, detected a signal of prolate rotation in the stellar field NGC1052-DF2 (see \citealt{2018Natur.555..629V,2019MNRAS.486.1192T,2019MNRAS.486.5670R} for a detailed discussion about the physical properties of this galaxy), \cite{2020MNRAS.491.3496C}  analyzed the And XIX extreme diffuse galaxy finding a marginal velocity gradient using $\sim 100$ giant branch stars, and \citet{2019ApJ...880...91V} used spatially resolved stellar kinematics of the  DF44 UDG to show that there is no evidence of rotation along its major axis.

In this manuscript we use  the NIHAO simulation suite in order to identify the  kinematical support of the stellar component of simulated field UDGs, and  we study the evolution of these systems as a function of redshift in order to identify the  mechanisms that induce  rotation versus  dispersion support in  UDGs.

This work is structured as follows: in section \ref{sec:sim}  we will explain the details of the NIHAO simulations  and the galaxy selection criteria; in section \ref{sec:RotOrPre} we will present the stellar kinematic maps of our simulated UDGs, determine a criterion to distinguish between rotation and dispersion support of their stellar component,  and show that the existence of two kinematically different populations can be related to the  gas accretion alignment across cosmic time. Finally in section \ref{sec:conl} we will make a summary of the results and present our main conclusions,  offering observational predictions for how many rotation vs. pressure supported UDGs are expected in the sky.

\section{Simulations}\label{sec:sim}

\subsection{NIHAO simulation suite}
\begin{figure*}
    \centering
    \includegraphics[width=1\linewidth]{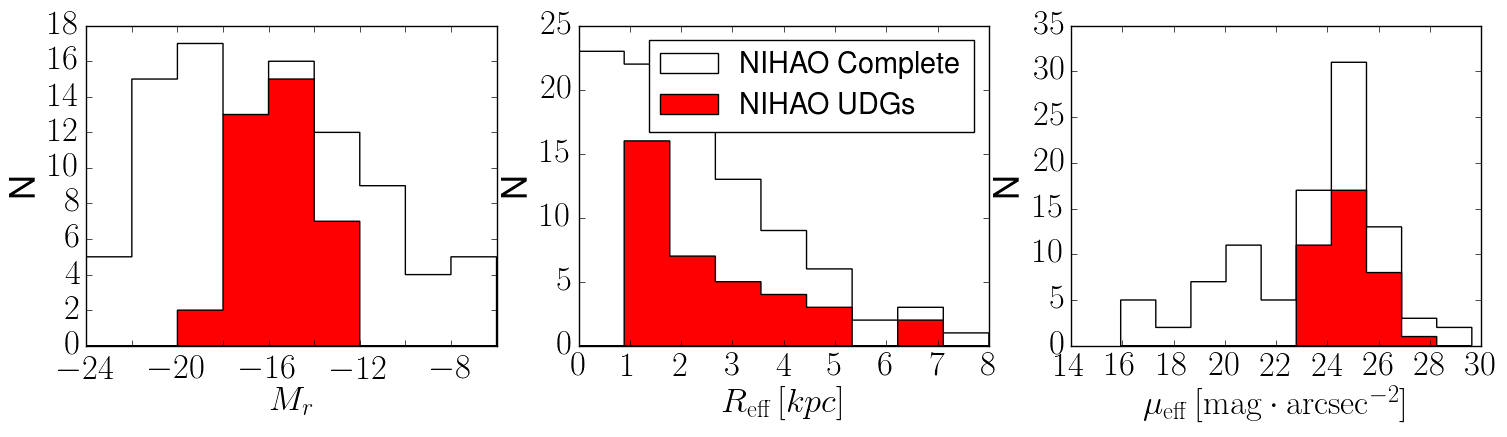}
    \caption{From left to right, we show the absolute magnitude in the $r$ band ($M_r$), the $2$D effective radius ($R_{\rm eff}$) and the effective surface brightness ($\mu_{\rm eff}$)  for all NIHAO galaxies (in white) and for our selected sample of UDGs (in  red). All properties have been computed in face-on configuration. In this work we define as UDG any galaxies with a ${\rm M}^{\ast}<10^9{\rm M}_{\odot}$ and  with a $R_{\rm eff}>1$kpc, which automatically  selects galaxies with a low surface brightness of $\mu_{\rm eff}>23.5 {\rm mag/arcsec}^2$: in doing so, we are including  some larger low surface brightness galaxies with respect to the sample presented in \citet{2017MNRAS.466L...1D}. We have a total of $37$ diffuse, low surface brightness objects satisfying our selection criteria, generally labeled as UDGs in this figure and through the paper.}
    \label{fig:criteria_hists}
\end{figure*}

We use a set of zoom-in simulated galaxies from the Numerical Investigation of a Hundred astrophysical Objects \citep[NIHAO project,][]{2015MNRAS.454...83W}. This project uses the N-body smooth particle hydrodynamics (SPH) code known as Gasoline2 \citep{2017MNRAS.471.2357W} and has been run using Planck Cosmology \citep{2014A&A...571A..16P}.

These simulations include three types of particles: dark matter, stars and gas. The way the gas particles form stars has been fixed in order to reproduce a Kenicutt-Schmidth law. Cool and dense enough gas particles, with density and temperature thresholds being $n_{th}>10.3 {\rm cm}^{-3}$ and $T<15000 K$ respectively, are eligible to form stars, and the cooling mechanisms include Hydrogen, Helium and Metal lines as well as Compton cooling \citep{2010MNRAS.407.1581S}.

Each star particle represents a population of stars. The number of stars of a given mass follows a Chabrier Initial Mass Function \citep{2003PASP..115..763C}. This type of IMF sets the number of massive stars formed in star formation events, which fixes the number of supernova (SN) events. These massive stars heat up the surrounding gas particles through stellar winds and radiation \citep{2006MNRAS.373.1074S}. This type of feedback is referred to as "early stellar feedback", and it has been shown to be essential in order to  reproduce galaxies that follow the correct scaling relations \citep{2012MNRAS.424.1275B,2013MNRAS.428..129S}. Later on, these massive stars explode as SN injecting energy and metals into the surrounding gas via shock waves as described in \cite{2006MNRAS.373.1074S}. In general, these SN events occur in high density environments where the details of the interstellar medium are not resolved, which leads to rapid over-cooling. In order to avoid this numerical issue, cooling is disabled within the blast-wave radius. 

Besides the heating through early stellar feedback and SNe feedback, the gas particles are also heated by a redshift dependent UV background \citep{2012ApJ...746..125H}.  

The metals injected to the ISM via SN (type II and type Ia) follow the classical yields from the literature (\citealt{1986A&A...158...17T} for SNe Ia and \citealt{1995ApJS..101..181W} for type II SNe) and their diffusion through the ISM is described in \cite{2010MNRAS.407.1581S}. 

For each galaxy the resolution has been set to resolve the mass profile down to the $1\%$ of the virial radius \citep{2015MNRAS.454...83W}, in order to resolve it inside the galaxy half -light radius, which is typically of that order of magnitude \cite[see][]{2013ApJ...764L..31K}. The galaxies from  NIHAO  are central isolated galaxies \citep[this is a requirement for the zoom-in technique, see][]{2015MNRAS.454...83W} that agree with the abundance matching predictions from \cite{2013MNRAS.428.3121M}.

 Since all NIHAO galaxies are isolated ones, the merger tree for each of them is straightforward to build, by following the most massive halo at any redshift. The halos at each snapshot where identified using Amiga Halo Finder \citep{2004MNRAS.351..399G,2009ApJS..182..608K}

\subsection{Galaxy selection}
\begin{figure}
    \centering
    \includegraphics[width=.49\textwidth]{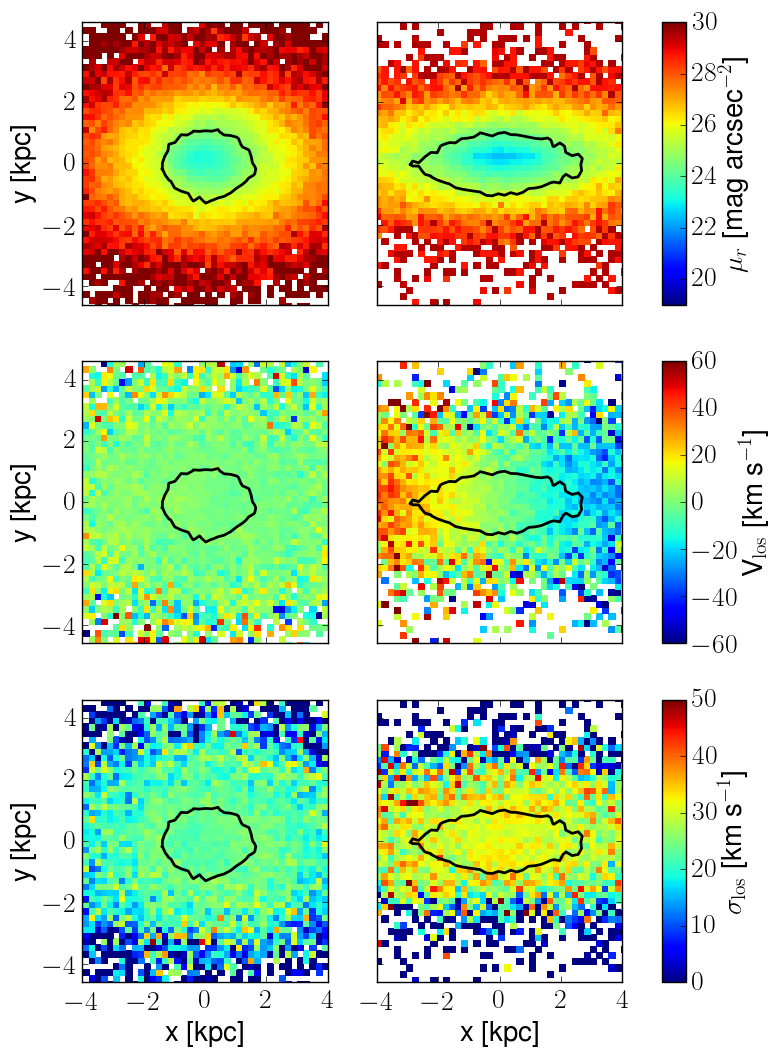}
    \caption{Stellar kinematic maps of two UDGs from the NIHAO simulation suite. From top to bottom we show respectively the surface brightness, line-of-sight stellar velocity and stellar velocity dispersion maps. The left column represents a clear case of dispersion supported UDG, while in the right column a strong  rotation component can be observed in the velocity field of the galaxy. Both objects are depicted  edge-on, with the black line representing the isophote that encloses the same area as a circle with radius equal to $R{\rm_{eff}}$: we will use these isophotes to compute the relative contribution to the stellar kinematic of rotation and dispersion, as shown in Fig 3.}
    \label{fig:maps}
\end{figure}

Our sample is a collection of simulated dwarf galaxies, all in isolation, with  stellar masses lower than $M_{\ast}$$<$$10^9{\rm M}_{\odot}$ and $2D$ effective radii larger than $R_{\rm eff}$$=$$1 {\rm kpc}$. We have checked that the selected galaxies are all well resolved, having at least $\sim9000$ stellar particles each.
With this selection the low-surface brightness nature of these galaxies is ensured; in practice, we are only selecting objects with effective surface brightness lower than $\mu_{\rm eff}$$\sim$$23.5 {\rm\, mag \, arcsec}^{-2}$, where $\mu_{\rm eff}$ is defined as :
\begin{equation}
    \mu_{\rm eff}[\rm mag\,  arcsec^{-2}] = \mathcal{M}_{\odot} + 21.572 - 2.5\log_{10}\left(\frac{(L/2)/L_{\odot}}{\pi (R\eff/pc )^2
    }\right)
\end{equation}
being $\mathcal{M}_{\odot}$ the sun's absolute magnitude; $L/2$ half the total luminosity of the galaxy (i.e. the luminosity enclosed inside an effective radius), and $R_{\rm eff}$ the $2D$ circularized effective radius. Both the effective radius and the effective surface brightness have been computed in $r$-band and in face-on projection\footnote{We define face-on projection as the inclination at which the vector of the stellar angular momentum is parallel to the line of sight.}.

From the full sample of NIHAO galaxies, $38$ objects meet these criteria, and we further excluded one galaxy which visually appears to be undergoing a merger at $z=0$.

We are therefore left with  a final sample of $37$ simulated low surface brightness, ultra-diffuse galaxies. Note that the selection criteria used in this work is a less restrictive version of the one used in \cite{2017MNRAS.466L...1D}, that followed the original UDG definition of \cite{2015ApJ...798L..45V}:   in  \cite{2017MNRAS.466L...1D} the authors selected galaxies only up to ${\rm M}_{\ast}=10^{8.5} {\rm M}_{\odot}$, therefore leading to a smaller sample than what reported here. 

In Fig. \ref{fig:criteria_hists} we show the distribution of magnitude, effective radii and effective surface brightness for the full NIHAO sample (white histograms) and for the UDG-like galaxies (red histograms).
\begin{figure*}
    \centering
    \includegraphics[width=1\textwidth]{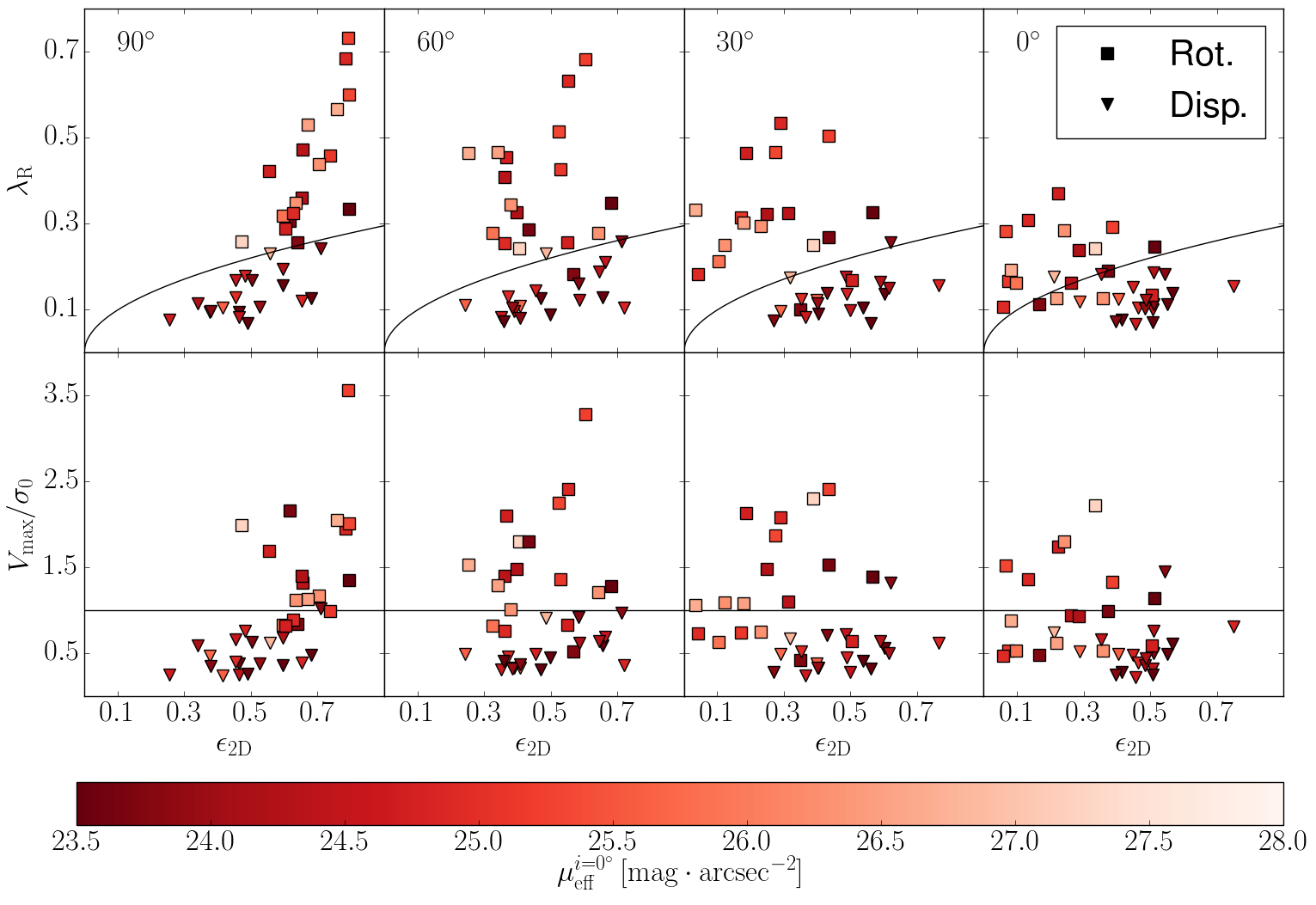}
    \caption{Characterization of the stellar kinematic of the simulated NIHAO UDG galaxies in our sample, through an anisotropy diagram ($\lambda_{\rm R}$, upper panels) and a $V_{\rm max}/\sigma_{0}$ diagram  (lower panels) vs. 2D ellipticity, for $4$ different inclinations of the galaxies (from left to right, $90\degr$, $60\degr$, $30\degr$ and $0\degr$). In every panels the UDGs are color-coded by their effective surface brightness, $\mu\eff$, in face-on configuration,  in order to ease the identification of galaxies in different projections. Here, $\lambda_{\rm R}$  is a proxy for the specific angular momentum of the stars \protect\citep[see][]{2007MNRAS.379..401E}, $V_{\rm max}$ is the maximum \textit{los}  velocity computed from the  kinematic maps, $\sigma_0$ is the line of sight velocity dispersion inside one effective radius and $\epsilon_{\rm 2D}$ is the projected ellipticity. These quantities  have been computed within the isophote described in  Fig. \ref{fig:maps} and in the text. The black solid line ($\lambda_{\rm R} = 0.31\sqrt{\epsilon_{2D}}$) splits rotation (above the line) vs dispersion (below the line) dominated galaxies, according to the \protect\cite{2011MNRAS.414..888E} definition. The markers indicate the galaxies that have been identified, in edge-on projection, as rotation (squares) or dispersion (triangles) supported, and we have maintained the same markers scheme for different inclinations in order to be able to track individual galaxies.
    About half of our simulated UDGs present a rotationally supported stellar kinematic, while the other half show a more dispersion supported stellar population.
    It can be observed that by moving from a edge-on (90$\degr$) to a face-on (0$\degr$) configuration, galaxies tend to move towards the bottom left side of the diagrams: that is, their stellar rotation decreases and so does their ellipticity, however  the  number of rotation supported galaxies decreases only slightly when observing the whole sample face-on (from 49$\%$ to 43$\%$ of the total).}
    \label{fig:rot_e}
\end{figure*}

\section{Rotation or pressure supported UDG$_{\rm s}$ }\label{sec:RotOrPre}
\subsection{Stellar kinematic Maps}
To characterize the stellar kinematic of our sample of UDGs, we have constructed maps of stellar flux, line-of-sight velocity and velocity dispersion for each galaxy at $10$ different projections, from edge-on view (i.e. stellar angular momentum  perpendicular to the \textit{los} velocity) to face-on view (i.e. stellar angular momentum  parallel to the \textit{los} velocity) in steps of $10\degr$. These maps have been constructed using an homogeneous grid with square bins of side of $0.2$ kpc.

The flux of each bin is the sum of the flux of all the star particles within it. The fluxes have been computed using the python module \texttt{pynbody}, which uses the Padova's library of isochrones of single stellar populations\footnote{http://stev.oapd.inaf.it/cgi-bin/cmd} inferred from  \cite{2010ApJ...724.1030G} and \citet{2008A&A...482..883M}. The line of sight velocity, $\overline{V_{los, i}}$, and velocity dispersion, $\sigma_{{ los}, i}$, are the mean and standard  deviation of the velocities of the stellar particles  within the bin:

\begin{equation}
\begin{aligned}
    \overline{V_{{ los}, i}} &= 
        \frac{1}{N_i}{\sum_{j=1}^{N_i} V_{{los},i}^{j}} 
    \hspace{.25cm};
    \hspace{.25cm}
    \sigma_{{los},i}=\sqrt{
        \frac{1}{N_i}\sum_{j=1}^{N_i} \left( V_{{ los},i}^{j} -  \overline{V_{{ los}, i}}\right)^2
        }
\end{aligned}
\end{equation}

\noindent where $N_{i}$ is the number of particles inside the $i$-th bin and $V_{{los}, i}^j$ is the line-of-sight velocity of the $j$-th particle in the $i$-th bin. 

In  Fig. \ref{fig:maps} we present an example of  maps constructed with this method for two of the UDG galaxies in our sample, shown edge-on. The black lines represent the isophote that encloses the same area as a circle with radius equal to R${\rm_{eff}}$: we will use the stars within such isophotes to compute the stellar kinematic properties.

It can be appreciated how the galaxy on the right shows a clear rotation component while the galaxy on the left is essentially dispersion-supported.
Indeed, as we will quantify in the next Section, within our simulated UDGs we found examples of both rotation as well as  dispersion supported systems.

\subsection{Rotation vs Dispersion support of the stellar component}

\begin{figure}
    \centering
    \includegraphics[width=.5\textwidth]{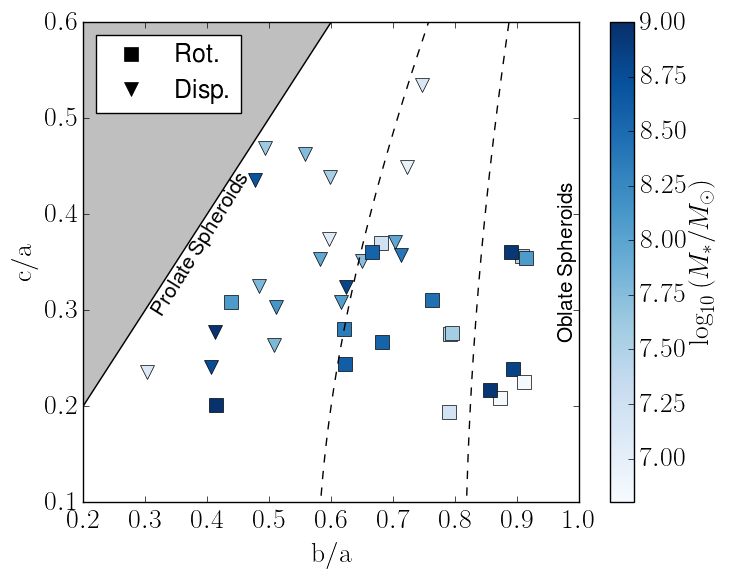}
    \caption{Short-to-long axes ratio (c/a) versus median-to-long axes ratio (b/a) for the stellar component of NIHAO UDGs. The grey zone shows the region in the plane where $c>b$ and cannot be populated by  definition ($a\geq b \geq c$). The dashed lines are,    from left to right, $T=1/3$ and $T=2/3$. Galaxies with $T$$\rightarrow$$1$ are oblate,  if    $T$$\rightarrow$$0$ they are prolate,  while the ones with intermediate values are triaxial \citep{2011ApJ...728..128D}. The markers represent rotation supported galaxies (squares) and dispersion supported galaxies (triangles) and the color code is their stellar mass.}
    \label{fig:triax}
\end{figure}

In order to quantify the rotation of the UDGs stellar component,  we use two parameters, namely $\lambda_{\rm R}$ and $V_{\rm max}/\sigma_0$.

The projected specific angular momentum, $\lambda_{\rm R}$, is defined as follows \citep[see][]{ 2007MNRAS.379..401E}:
\begin{equation}
    \lambda_{\rm R} = \frac{\left<r_{\perp}|V_{ los}|\right>}{\left<r_{\perp}\sqrt{V_{los}^2 +\sigma_{los}^2}\right>}
\end{equation}

where $r_{\perp}$ is the projected distance of each bin to the galaxy center and  $\left<\right>$ denotes the flux weighted average over the galaxy map. The dependence of $\lambda_{\rm R}$ from the physical distance to the center makes this parameter dependent on the aperture used:  we follow the prescription of \cite{2007MNRAS.379..418C} and  limit the summation to the isophote that encloses the same area as a circle with radius equal to the effective radius, i.e. $A_{\rm isophote} = \pi R_{\rm eff}^2$.

Furthermore, we also  use the classical $V_{\rm max}/\sigma_0$ criteria for rotation, where  $\sigma_0$ is the \textit{los} velocity dispersion inside a circular aperture of half effective radius \citep[see][]{2005MNRAS.363..937B} and $V_{\rm max}$ is the maximum \textit{los} velocity inside the isophote defined above. We decided  to use this aperture limit to compute  $V_{\rm max}$ in order to avoid more external bins with less star particles, insufficient statistics and therefore unrealistic values of $V_{\rm max}$ . 

It is common to study the rotation support of a galaxy as a function of its shape. For this, we have computed the mean $2D$ ellipticity as proposed by \cite{2007MNRAS.379..418C}:
\begin{equation}
    \epsilon_{2D} = 1 - \sqrt{\frac{\sum_iF_iy^2_i}{\sum_iF_ix^2_i}}
\end{equation}

where $F_i$, $y_i$, $x_i$ are the flux and the distance of the $i$-th bin to the galactic center along the minor and major axes respectively, and the calculation has been performed using only particles within the isophote defined above. 

In order to distinguish between rotation and dispersion supported systems we have followed \cite{2011MNRAS.414..888E}, being those galaxies with $\lambda_{\rm R}$$>$$ 0.31\sqrt{\epsilon_{2D}}$ rotation supported systems, while those with lower $\lambda_{\rm R}$$<$$ 0.31\sqrt{\epsilon_{2D}}$ dispersion supported ones. Since this criterion has been calibrated for more massive galaxies, we have compared such a  selection with a visual inspection of the kinematic maps of our galaxies, finding a reasonable match.  

In Fig. \ref{fig:rot_e} we show the $\lambda_{\rm R}-\epsilon_{2D}$ and $V_{\rm max}/\sigma_{0}-\epsilon_{2D}$ planes for our sample of UDGs at $4$ different projections, from edge-on view to face-on view (left to right panels), color coded by the effective surface brightness in face-on configuration.
The black solid lines ($\lambda_{\rm R} = 0.31\sqrt{\epsilon_{2D}}$) split rotation (above the line) vs dispersion (below the line) dominated galaxies, although we note that our simulated UDGs follow a continuous distribution  from   dispersion to  rotation supported systems, rather  than showing two distinct populations.

As it can be seen in Fig. \ref{fig:rot_e} comparing the top and bottom panels, both methods provide similar results: a high value of $\lambda_{\rm R}$  corresponds to a high value of $V_{\rm max}/\sigma_{0}$, and the evident correlation between the amount of rotation and the ellipticity  of a galaxy vanishes when we  move from edge-on to face-on projection. 

The most reliable measurement of the galaxy rotation is the one made in edge-on projection:  using this configuration  as a proxy for the intrinsic rotation, we found that $18$ out of the $37$ simulated UDGs are rotation supported systems ($\sim 49\%$). Rotation supported UDGs seem to be disky-like galaxies with large ellipticities when seen edge-on ($\overline{\epsilon_{2D}} \sim 0.7$) but much lower $\epsilon_{2D}$ when seen face-on ($\overline{\epsilon_{2D}}\sim 0.2$).

\begin{figure}
	\centering
	\includegraphics[width=1\columnwidth]{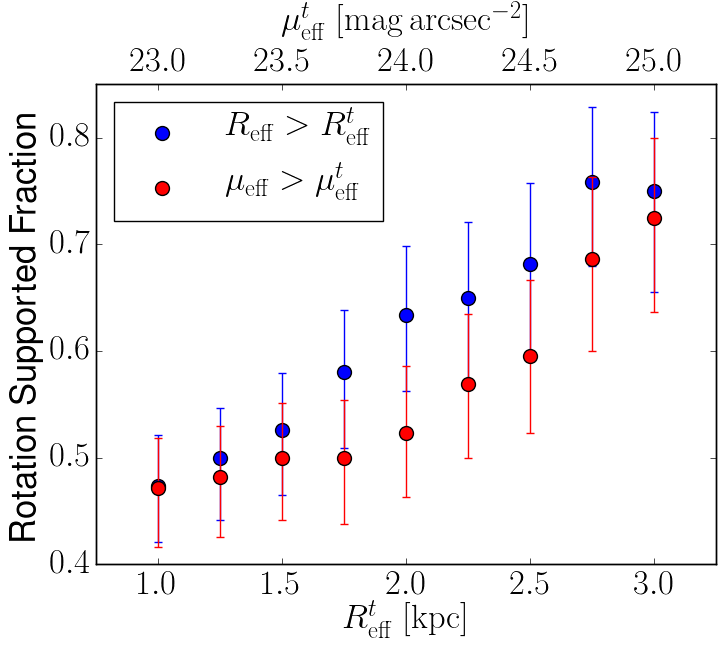}
	\caption{ Variation of the measured rotation supported fraction of simulated field UDGs as a function of the selection criteria used to define UDGs. From our sample of $370$ galaxies ($37$ with $10$ different projections each) we compute the rotation fraction assuming different  selection criteria. We explore the effects of filtering the sample using different thresholds in effective radius ($R_{\rm eff} > R_{\rm eff}^t$, in blue) and effective surface brightness ($\mu_{\rm eff} > \mu_{\rm eff}^t$, in red). For each resulting sample we have computed the rotation supported fraction distribution with a resampling method described in the text. The markers are the median of the distribution and the errorbars correspond to the $16th$ and $84th$ percentiles. } 
	\label{fig:RotFrac}
\end{figure}

\begin{figure*}
    \centering
    \includegraphics[width=1\textwidth]{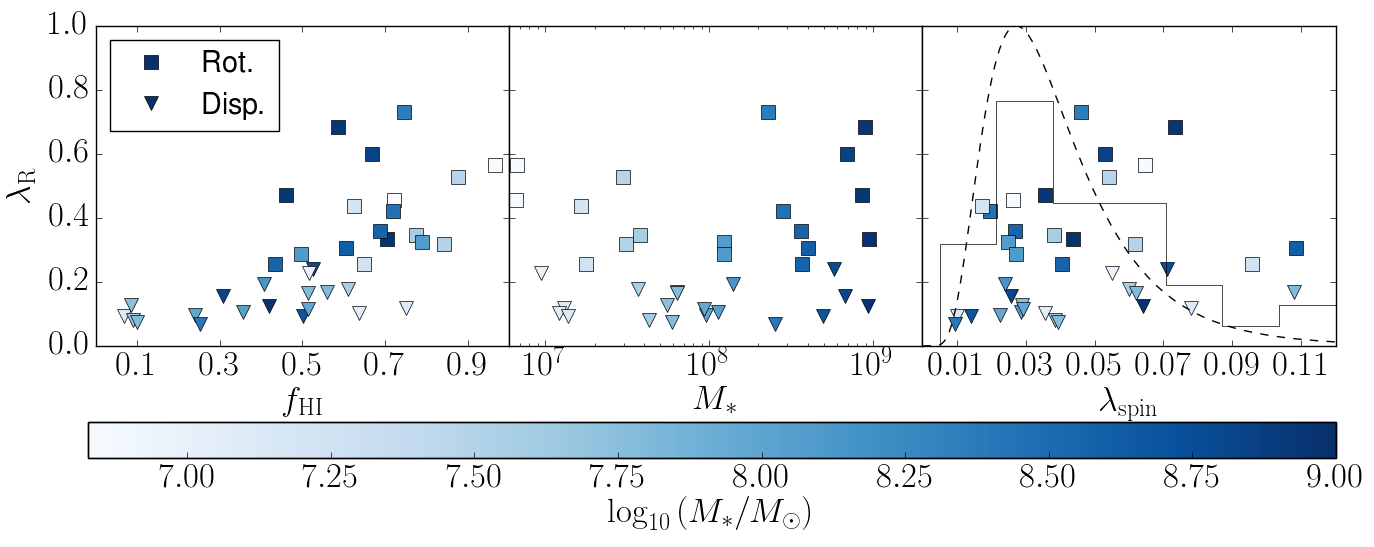}
    \caption{The rotation parameter $\lambda_{\rm R}$ as a function of HI gas fraction ($f_{\rm HI}$), stellar mass ($M_{\ast}$) and dark matter halo spin ($\lambda_{\rm spin}$) for our sample of UDGs, color coded  by their stellar mass. We indicated as squares the rotation supported galaxies (those for which $\lambda_{\rm R}> 0.31\sqrt{\epsilon_{2D}}$, when computed edge-on), while the  dispersion supported systems are shown as downwards triangles. The UDGs HI fractions correlate with the rotation support. No correlation is observed between the rotation  of the  UDGs' stellar  component and their stellar mass  or  halo spin. Moreover, we show that the distribution of spins of UDGs is quite standard, by plotting, in the last panel, a normalized histogram of UDGs' spins (solid line) and the expected spin distribution from \protect\cite{2001ApJ...555..240B} (dashed line).}
    \label{fig:HI}
\end{figure*}

Regarding dispersion supported UDGs, we note that their ellipticity, which is always different form zero, appears to be more or less constant with the inclination: this can only be explained if those galaxies are triaxial or even prolate spheroids.

We have checked this by computing the $3D$ axes ratios of our sample, shown in  Fig. \ref{fig:triax}. There, dispersion supported galaxies are indicated as triangles while rotation supported ones as squares, and the color code refers to their stellar mass. 
We start by selecting a spherical aperture of radius equal to the $3D$ half light radius of the galaxy and computing the flux weighted inertia tensor of the stars; we then use such eigenvalues to define new axes ratios and we keep iterating until the variation in the semi-axes is smaller than $1$ pc.
The dashed lines show fixed values of the triaxiality parameter,  defined as $T=\left[1 - (b/a)^2\right]/\left[1-(c/a)^2\right]$, such that from left to right we move from prolate to oblate objects, while in-between values are completely triaxial  galaxies.

In the dwarf regime the galaxies tend to be triaxial \citep{2019RNAAS...3..191S,2019ApJ...883...10P,2013MNRAS.436L.104R} and, as can be seen  in Fig. \ref{fig:triax}, UDGs are not an exception. In general, Fig. \ref{fig:triax} highlights the relation between  the  morphologies of these galaxies and the kinematical support of their stellar population, being the dispersion supported UDGs very triaxial or prolate spheroids, while rotation supported UDGs are more oblate galaxies.

As we stated before, with our selection criteria, half of our UDGs ($18$ out of $37$ ) can be classified as intrinsic rotation supported (i.e. $\lambda_{\rm R} > 0.31\sqrt{e_{\rm 2D}}$ in edge-on). We would now like to compute the expected fraction of rotation (or dispersion) supported UDGs assuming random orientations. To do so, we firstly  generated $10$ inclinations for each of our  $37$ galaxies, in order to obtain $370$ galaxies with different viewing angles. We have taken into account the fact that because of projection effects some UDGs do not meet our selection criteria (i.e. $R_{\rm eff} > 1 {\rm kpc}$). From this sample, we then  randomly selected $100$ galaxies and compute the fraction of rotation vs. dispersion supported objects, and we repeated the procedure  $1000$ times. Thus, if the NIHAO UDGs are a representative sample of the population of field UDGs we predict that an observational survey will find a fraction of $47\pm5\%$ rotationally supported UDGs. 

 We are also interested in exploring how different selection
criteria will affect this fraction. As is shown in Fig \ref{fig:RotFrac}, selecting more diffuse galaxies, either filtering by larger effective radius (e.g.  \citealt{2015ApJ...807L...2K} select galaxies with $R_{\rm eff} > 1.5 $ kpc) or by higher surface brightness, will bias the sample towards larger fraction of rotationally supported UDGs.  As an example, using in our sample a selection criteria of $R_{\rm eff}>2.5$ kpc ($\mu_{\rm eff} > 24.5 $ mag arcsec$^{-2}$) will lead to a rotation supported fraction of $68\pm8\%$ ($60\pm7\%$).
In Fig. \ref{fig:RotFrac} the point that corresponds to our selection criteria is $R_{\rm eff} >1$ kpc which gives a value of $47\pm5\%$ fraction of rotation supported galaxies.

\subsection{Gas fraction and  halo spin}
In this section we are going to explore the difference in the physical properties of  UDGs as a function of their kinematical support;  
in particular, we will study their HI gas fraction, stellar mass and halo spin, to investigate how these quantities correlate with the $\lambda_R$ parameter.

\begin{figure}
    \centering
    \includegraphics[width=.5\textwidth]{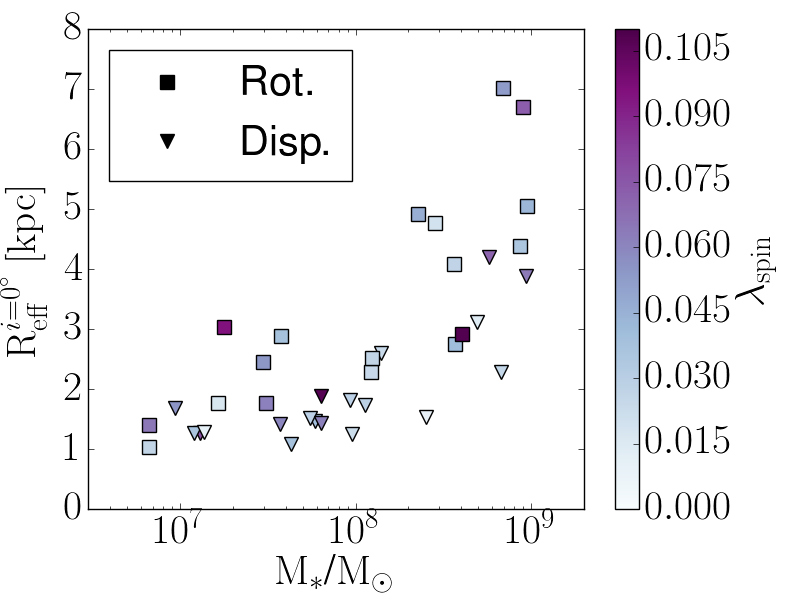}
    \caption{Face-on effective radius as a function of  stellar mass for rotation supported (squares) and dispersion supported (triangles) UDGs, color coded by their dark matter halo spin. As expected, at a fixed stellar mass, rotation supported UDGs have more extended stellar distributions (larger R$_{\rm eff}$). Moreover, we found no correlation between effective radius and $\lambda_{\rm spin}$.}
    \label{fig:ReffMs}
\end{figure}

\begin{figure*}
    \centering
    \includegraphics[width=1\textwidth]{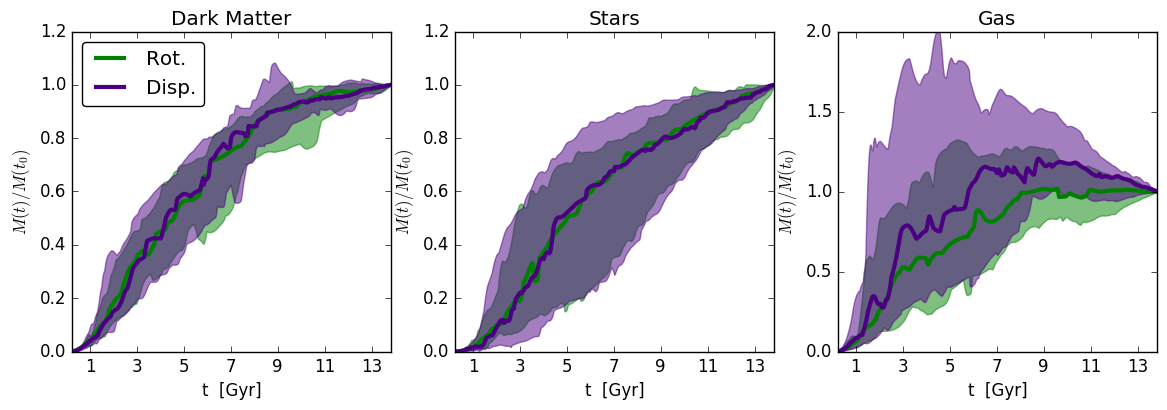}
    \caption{Median mass accretion rate for  rotation (green) and dispersion (violet) UDGs. From left to right we show the dark matter,  stars and  gas accretion through cosmic times, normalized to the relative mass value at z=0. The shaded regions represent the $16th$ and $84th$ percentiles of the population. No difference appears in the average evolution of the dark matter and of the  stellar component, while the gas is accreted faster in the dispersion supported galaxies (albeit with a large associated dispersion over the average).
     The different stellar kinematics of UDGs could therefore depend on the way that such gas is accreted, as explored in Fig. \ref{fig:Alinemaineto},
     such that galaxies with earlier inflows and stronger build-up of gas over a short period of time  will likely end up being dispersion supported.
}
    \label{fig:mass}
\end{figure*}

In the left panel of  Fig. \ref{fig:HI} we present the rotation parameter $\lambda_R$ vs the HI gas fraction, $f_{\rm HI} = \rm M_{\rm HI}/(\rm M_{\rm _{HI}} + \rm M_{\ast})$, for dispersion supported (inverted triangles) and rotation supported (squares) UDGs. While it was already shown in \cite{2017MNRAS.466L...1D}  that NIHAO UDGs are on average HI gas rich galaxies, here we  further demonstrate that rotation supported UDGs are the HI rich-most galaxies, with a median HI fraction of $\sim 0.7$, while dispersion supported UDGs, despite having on average $f_{\rm HI} \sim 0.4$, present a tail in HI fraction distribution that extends towards HI poor galaxies.

In the central panel of Fig. \ref{fig:HI} we explore the possibility that the stellar kinematical properties of  UDGs depend on their stellar mass: we found no appreciable differences between the two populations, being both rotation and dispersion supported UDGs present at every stellar mass considered here. 
Nevertheless we note, as expected, that at fixed  stellar mass rotation supported galaxies have a larger effective radius than dispersion supported ones, as depicted in Fig. \ref{fig:ReffMs}. 
On the other hand, this plot also explains the behaviour shown in Fig \ref{fig:RotFrac}. When  using a larger effective radius as a selection criteria, we are removing a large fraction of dispersion supported galaxies (independently of their inclination) and we are left with intrinsically rotation supported ones.

We remind that, as  shown in \cite{2014MNRAS.437..415D},  the stellar mass range  (10$^7$$<$$\rm M_*/\rm M_{\odot}$$<$$10^9$) is where SN-driven core formation processes are most efficient, and that these processes of expansion  have been shown to be a viable formation mechanism for forming field UDGs. However, it seems that the contribution of SN feedback plays only a secondary role in 
 generating differences in the stellar kinematic distribution of rotation and dispersion supported  UDGs. 
Despite the fact that \textit{all} of these galaxies undergo through strong supernova feedback, some of them end up being dispersion and some other rotation supported systems, and we will explore in the next Section which is the mechanism responsible for this.

Finally, we  explored the possibility that rotationally supported galaxies live in high-spin haloes, by representing the   $\lambda_{\rm R}$ parameter as a function of the dark matter halo spin $\lambda_{\rm spin}$ \citep{2001ApJ...555..240B}, in the right-side panel of Fig. \ref{fig:HI}. The $\lambda_{\rm spin}$ has been obtained using the equivalent dark matter only simulations.  We see no differences between the rotation supported galaxies and dispersion supported ones in their halo spin distribution.  Moreover, for both populations, the spin distribution follows the expected distribution from $\Lambda$CDM cosmology 
\footnote{We verified this by  performing a Kolmogorov-Smirnov test.}, 
shown as solid and dashed histograms, respectively:  this means that our UDGs do not live into high-spin haloes, as proposed by \cite{2016MNRAS.459L..51A}, even thought this seems to be the case for  more massive low-surface brightness galaxies, which show a clear correlation between their $\mu_{\rm eff}$ and $\lambda_{\rm spin}$ \citep[see][]{2019MNRAS.486.2535D}.

\subsection{Origin of the two kinematically distinct populations}
\label{sec:causses}

 In this section we explore the evolution of NIHAO UDGs  to determine the conditions that lead to their different stellar kinematics.

First we focus on the galaxies' mass accretion history in  Fig.~\ref{fig:mass}, where we show, from left to right, the dark matter halo, stellar mass and gas mass accretion over cosmic time. We indicated rotation supported systems in green and dispersion supported ones in violet, and show the normalized mass accretion to its z=0 value (i.e. ${\rm M}(t)/{\rm M}(t_0)$).
As shown in the two left-most panels of Fig. \ref{fig:mass},  there is no difference between the dark matter and the stellar mass accretion histories of the two populations: both rotation and dispersion supported systems form at a similar time, specifically their half mass formation time is about t$\sim$4 Gyrs for their dark matter haloes and t$\sim$5 Gyrs for their stellar mass.

Nevertheless, it can be appreciated in the gas accretion history (right panel of Fig. \ref{fig:mass}) that the dispersion supported population acquires on average its gas mass earlier than the rotation supported population, and that by t$\sim$6 Gyrs its total gas mass has already reached its value at z=0: that is, the dispersion supported UDGs accrete gas mass fast  and they then decrease it via strong outflows driven by SNae feedback. 
The large variation in the gas accretion history of the dispersion supported population is dominated by the less HI-gas rich galaxies ($f_{\rm HI} < 0.3$): these galaxies have early gas injections, a consequently initial burst of star formation, and then evolve passively loosing its gas due to strong stellar feedback. 

We found that rotation supported UDGs retain a factor of 1.5 to 2 times more baryons than dispersion supported ones, taking into account the dependence of baryon fraction on total mass. Moreover, the gas in the dispersion supported UDGs is mostly in  hot-warm phase: indeed,  we  verified that in the dispersion supported cases, the temperature of the gas is hotter than the one of the rotationally supported galaxies at z=0, by about $5000$ K at $\sim$10 kpc (and as we go to outer radii the temperature difference increases even more), a sign of the chaotic state of the gas in these systems. In addition, as we saw in Fig \ref{fig:HI}, dispersion supported UDGs  have  almost an order of magnitude less HI gas than rotation ones. We summarize the results of average baryonic, HI and total mass in  Table \ref{tab:Masses}.

\begin{table}
    \centering
    \begin{tabular}{c|cc}
         \hline
                                & Rotation Supported & Dispersion Supported \\ \hline
         $\left<\rm M_{\rm DM}/M_{\odot}\right>$  & $6.43\cdot10^{10}$  & $4.18\cdot10^{10}$ \\
         $\left<\rm M_{\rm b}/M_{\odot}\right>$ & $4.13\cdot10^9$ & $1.79\cdot10^9$ \\
         $\left<\rm M_{\rm gas}/M_{\odot}\right>$ & $3.88\cdot10^9$     & $1.67\cdot10^9$ \\
         $\left<\rm M_{\rm HI}/M_{\odot}\right>$  & $3.72\cdot10^{8}$   & $6.33\cdot10^7$ \\
         \hline
    \end{tabular}
    \caption{Average $z$=0 masses for the two samples of UDGs: rotation supported (first column) and dispersion supported (second column). In order, from top to bottom, dark matter, baryonic, total gas and HI gas mass.}
    \label{tab:Masses}
\end{table}

To further investigate the link between gas accretion and the final kinematic state of the stars, we  explore   the alignment of such accretion. For more massive systems it has already been shown that the orientation of the baryons at early times tightly correlates with the morphology of the galaxy at $z$=0 (see \citealt{2012MNRAS.423.1544S} and references therein).
Here, we  followed the same procedure described in \cite{2019MNRAS.486.2535D}:  we first identify all the cold gas and star particles that belong to the galaxy at $z=0$, we then trace them back to the redshift of half stellar formation time of the galaxy,  and we finally compute the angular momentum of this baryonic component in spherical shells that enclose different percentages of the baryonic mass ($\Vec{L_{f}}$, where $f$ is the percent of the mass).
The largest shell we have considered  encloses the $95\%$ of the tracked-back baryonic mass. 
The alignment is then defined as the angle $\theta$ between the angular momentum of each shell and the angular momentum of the most massive shell, at the galaxy's half-stellar mass formation time:

\begin{equation}
   \cos{\theta} = {\Vec{L_{f}}\cdot\Vec{L_{95}}}/\left({|\Vec{L_{f}}||\Vec{L_{95}}|}\right)
\end{equation}

\begin{figure*}
    \centering
    \includegraphics[width=.5\linewidth]{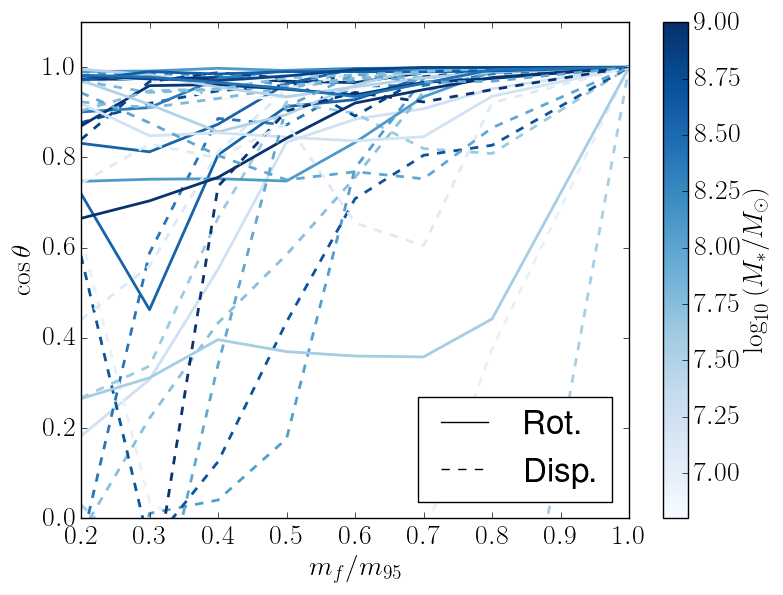}%
    \includegraphics[width=.48\linewidth]{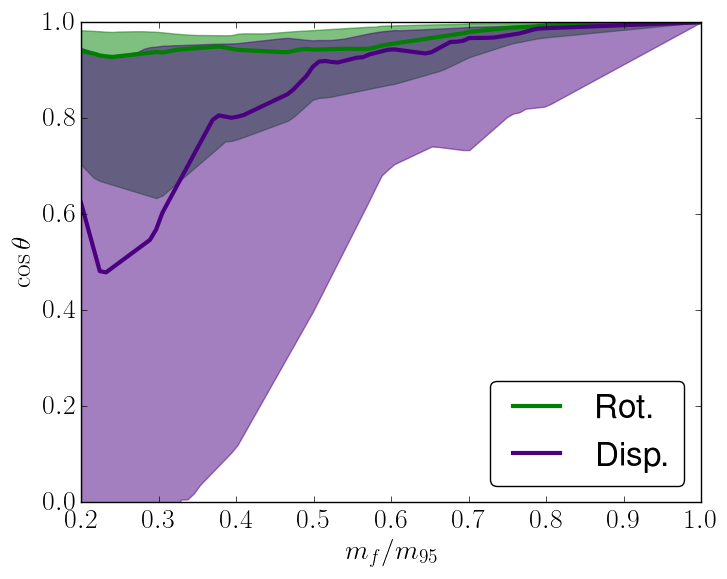}
    \caption{Alignment of the baryons infalling into the proto-galaxy  at half   its stellar formation time for NIHAO UDGs, for each galaxy individually (left panel, in which dashed lines  represent  pressure supported UDGs and  solid lines rotation supported ones) and as average for the two kinematically distinct groups (right panel, in which the dispersion  supported group is indicated in violet  and the rotation one  in green). We show the  angle $\theta$ between the angular momenta of inner and outer shells of infalling baryonic material vs the ratio m$_{\rm f}$/m$_{95}$, where m$_{95}$ is the largest shell considered, which includes 95$\%$ of all baryons that will belong to the galaxy by z=0, and m$_{\rm f}$ is a shell containing f $\%$ of the total baryonic mass. Lines that are close to unity for every value of m$_{\rm f}$/m$_{95}$ represent UDGs whose infalling baryonic material is accreting in a ordered and  aligned manner: this is the case for rotation supported UDGs.}
    \label{fig:Alinemaineto}
\end{figure*}

In Fig. \ref{fig:Alinemaineto} we show the alignment of the different accreting shells for all our  NIHAO UDGs. We can clearly see that the rotation supported UDGs have their infalling baryons already well aligned with the angular momentum of the proto-galaxy at half stellar mass formation time, which means that during the accretion process the baryons keep adding angular momentum to the proto-galaxy, eventually forming the rotation supported stellar component. The dispersion supported UDGs, instead, experience a more chaotic gas accretion: such mis-aligned gas build-up will lead to a dispersion supported stellar system.  

We would like to remark that both kinematic groups experience strong SNae feedback, which is indeed the main formation mechanism  for UDGs in our simulations;  the signatures of aligned vs. chaotic gas accretion at early times, however, are still visible in the UDGs stellar kinematics today. 
In other words, the memory of the past ordered gas accretion is retained in nowadays stellar kinematic, implying  that in general SNae feedback does not destroy the rotation acquired via aligned accretion of baryons.
 
\section{Conclusions}
\label{sec:conl}

 In this work we have studied a simulated sample of Ultra Diffuse Galaxies (UDGs) from the NIHAO project with stellar masses between $10^{6.5} < {\rm M}_{\ast}/{\rm M}_{\odot} < 10^{9}$. This set of zoom-in simulations have already been shown to be able to create UDGs via supernova-feedback-driven gas outflows \citep[e.g.][]{2017MNRAS.466L...1D} in galaxies with stellar mass lower than $<10^{8.5}{\rm M}_{\ast}$.
 In order to study the stellar kinematics of these systems we have constructed  line-of-sight stellar velocity and velocity dispersion maps (Fig. \ref{fig:maps}), and  used the projected specific angular momentum of their stellar component, $\lambda_{\rm R}$, to quantify their kinematical support:  we found that our simulated UDGs  are continuously distributed from dispersion to rotation supported systems (Fig. \ref{fig:rot_e}).  We made a comprehensive study of the evolution of these systems to understand whether the different stellar kinematics arise from different evolutionary paths, and in doing so we defined a UDG to be rotationally supported if $\lambda_{\rm R}$$>$$ 0.31\sqrt{\epsilon_{2D}}$, where $\epsilon_{2D}$ is the galaxy ellipticity (see also \citealt{2011MNRAS.414..888E}).\\
 
 The results of this paper can be summarized  as follows:
 \begin{itemize}
     \item nearly half of our isolated simulated UDGs ($\sim$ $49\%$) has a rotationally supported stellar disk, while the remaining  UDGs show a dispersion dominated stellar component (Fig. \ref{fig:rot_e});
     \item the morphology of these galaxies is related to the kinematical support of their stellar population:  the dispersion supported UDGs are triaxial or prolate spheroids, while rotation supported UDGs are more oblate (Fig. \ref{fig:triax});
     \item accounting for  random inclinations effects,  we expect that a future survey of field UDGs would be able to find about $47\pm5\%$ of these galaxies  to be rotation supported  when selecting them by effective radius larger than $1$ kpc, while more restrictive selection criteria will lead to larger fractions of rotation supported UDGs (Fig. \ref{fig:RotFrac});
     \item both dispersion and rotation supported UDGs live into dark matter haloes with standard halo spin parameter $\lambda_{\rm spin}$ that follows $\Lambda$CDM cosmology (Fig. \ref{fig:HI}), and underwent  a similar halo mass accretion history (Fig. \ref{fig:mass});
     \item  on average, simulated UDGs are HI rich galaxies, whose HI gas fraction correlates with their stellar kinematics, being the rotation supported UDGs the HI richest galaxies (Fig. \ref{fig:HI});
     \item  dispersion supported UDGs have, by z=0, a factor of $\sim$2 lower baryon fraction and hotter gas within their virial radius, than rotation supported ones (Table \ref{tab:Masses});
     \item the alignment of the infalling baryons at early times plays a fundamental role in determining the final kinematics of the stellar component of UDGs at z=0: galaxies that accrete their gas in an ordered manner will end up having rotation supported stellar disks, while mis-aligned accretion will result in dispersion supported UDGs (Fig. \ref{fig:Alinemaineto}). 
 \end{itemize}
 
 The importance of the alignment of accreted gas in determining the formation of a rotating disk is in agreement with previous results that focused on more massive galaxies (see \citealt{2012MNRAS.423.1544S,2019MNRAS.486.2535D} and references therein).

We remark that all our UDGs in the stellar mass range of  $10^{6.5} < {\rm M}_{\ast}/{\rm M}_{\odot} < 10^{9}$ go through strong outflows via SN feedback, which is the main formation mechanism of UDGs in our simulations, as presented in \cite{2017MNRAS.466L...1D}. 
However ,we should also note that some UDGs approach the transition regime (${\rm M}_{\ast}\sim10^{9}{\rm M}_{\odot}$) between feedback-dominated and angular-momentum-dominated formation scenarios for NIHAO low surface brightnes galaxies \citep[see][]{2019MNRAS.486.2535D}. Indeed, about $1/3$ of the UDGs presented in the current manuscript are in between the two ranges, such that the combined effect of SN feedback and aligned acccretion of gas can be expected to play a substantial role in forming them. In any case we show here that feedback is not strong enough to destroy the ordered rotation of the stellar component, which indeed retains its memory of the way the baryons have fallen into the protogalaxy.
The aligned accretion of baryons at early times translates into an  additional  angular momentum in the baryonic component, which give rise to even larger stellar distributions (Fig. \ref{fig:ReffMs}) and defines the resulting morphology of the galaxy (Fig. \ref{fig:triax}).

\section{Acknowledgements}
We thank the anonymous referee   for their useful report. This research was carried out on the High Performance Computing resources at New York University Abu Dhabi (UAE) and at the LaPalma supercomputer  (Spain).
We made use of the programming language \texttt{Python}\footnote{https://www.python.org/}. Data analysis was partially performed using the  module \texttt{pynbody} \citep{pynbody}. 
We thank Ignacio Trujillo for fruitful discussions.
SCB is supported by a FPI MINECO studentship. 
ADC acknowledges financial support from a MSCIF grant, H2020-MSCA-IF-2016 Grant agreement 748213, DIGESTIVO.
CBB thanks MINECO/FEDER grant AYA2015-63810-P and the Ramon y Cajal fellowship program. 
JFB and MAB acknowledges support through the RAVET project by the grant AYA2016-77237-C3-1-P from the Spanish MCIU.
MAB gratefully acknowledges support from the Severo Ochoa excellence pro-gramme (SEV-2015-0548)
TRL acknowledges financial support through the grants (AEI/FEDER,UE) AYA2017-89076-P, AYA2016-77237-C3-1-P, and AYA2015-63810-P, as well as MCIU, through the State Budget, and MCIU Juan de la Cierva - Formaci\'on grant (FJCI-2016-30342). 
JFB, TRL and MAB are grateful to the Consejer\'\i a de Econom\'\i a, Industria, Comercio y Conocimiento of the Canary Islands Autonomous Community, through the Regional Budget (including IAC project TRACES). 

\section{DATA AVAILABILITY}
The data underlying this article will be shared on reasonable request to the corresponding author.


\bibliographystyle{mnras}
\bibliography{biblio.bib}


\bsp	
\label{lastpage}
\end{document}